\documentclass[dvips]{article}

\usepackage{icrc2011}

\title{Multi-Wavelength Observations of the HBL Object 1ES 1011+496 in Spring 2008}
\newcommand{\etal}{\MakeLowercase{\textit{et al. }}} % "et al."
\shorttitle{Reinthal \etal Multi-Wavelength observations of 1ES 1011+496 in Spring 2008}

\authors{Riho Reinthal$^{1}$, Stefan R\"ugamer$^{2}$, Elina J. Lindfors$^{1}$ , Daniel Mazin$^{3}$, Antonio Stamerra$^{4}$ for the MAGIC collaboration, Francesco Longo$^{5}$, Fabrizio Lucarelli$^{6}$, Carlotta Pittori$^{6}$ for the AGILE team and Anne L\"ahteenm\"aki$^{7}$}
\afiliations{$^1$Tuorla Observatory, University of Turku, FI-21500 Piikki\"o, Finland\\ $^2$Universit\"at W\"urzburg, D-97074 W\"urzburg, Germany \\$^3$Institut de Fisica d'Altes Energies (IFAE), E-08193 Bellaterra (Barcelona), Spain \\ $^4$Universit$\grave a$ di Siena, and INFN Pisa, I-53100 Siena, Italy \\$^5$Universit$\grave a$ di Trieste and INFN Trieste, I-34127 Trieste, Italy \\$^6$Agenzia Spaziale Italiana (ASI) Science Data Center, I-00044, Frascati, Italy \\$^7$Aalto University Mets\"ahovi Radio Observatory, FI-02540 Kylm\"al\"a, Finland }
\email{rirein@utu.fi}

\abstract{In the spring of 2008 MAGIC organised multi-wavelength (MWL) observations of the blazar 1ES 1011+496. 1ES 1011+496 is a high-frequency peaked BL Lac object discovered at VHE $\gamma$-rays by MAGIC in spring 2007 during an optical outburst reported by the Tuorla Blazar Monitoring Programme. MAGIC re-observed the source during the 2008 MWL campaign which also included the Mets\"ahovi, KVA, Swift and AGILE telescopes. This was the first MWL campaign on this source that also included VHE coverage. MAGIC observed 1ES 1011+496 from March 4th to May 24th 2008 for a total of 27.9 hours, of which ~20 h remained after quality cuts. The observations resulted in a detection of the source at $\sim$7 $\sigma$ significance level with a mean flux and spectral index similar to those during the discovery.

Here we will present the results of the MAGIC observations of the source in combination with contemporaneous observations at other wavelengths (radio, optical, X-rays, high energy $\gamma$-rays) and discuss their implications on the modelling of the spectral energy distribution. }
\keywords{AGNs, BL Lac objects, 1ES 1011+964, multi-wavelength campaigns, VHE $\gamma$-rays, MAGIC }

\begin{document}
\maketitle

%Begin the section.
\section{Introduction}

Active Galactic Nuclei (AGN) are an extreme class of galaxies in which matter falling into the central black hole (BH) causes the nucleus to have much higher than normal luminosity over most, if not all of the electromagnetic spectrum. At least some AGNs develop jets of relativistic particles -- particles that travel at nearly the speed of light -- that shoot out from close to the poles of the central BH. In the case of blazars, a subclass of AGNs, these jets are oriented close to our line of sight, which causes an apparent increase in the jet luminosity and a decrease in their variability time scales across all wavelengths due to relativistic beaming effects.

The emission from AGNs and consequently also blazars is dominated by non-thermal radiation. Their spectral energy distribution (SED) is characterised by two broad peaks of which the lower energy one is believed to originate from synchrotron emission of charged particles in the jet. The higher energy peak is most commonly explained by inverse Compton scattering of either the synchrotron (synchrotron self Compton - SSC, see e.g. \cite{maraschi92, costamante02}) or external (external Compton - EC, \cite{dermer93, ghisellini05}) seed photons by the electrons and positrons in the jet, but hadronic models \cite{mannheim93, mücke03} are also considered. Blazars are divided into flat spectrum radio quasars and BL Lacertae objects. The latter are further subdivided into low-, intermediate- and high-frequency peaked BL Lac (LBL, IBL and HBL, respectively) objects according to the frequency of the first peak which in the case of HBLs is located in the UV to soft X-ray regime. The second peak in HBLs is located in the GeV to TeV range which makes them good candidates to detect very high energy (VHE, $>$100 GeV) $\gamma$-rays from.

Blazars show variability in flux at all wavelengths as well as spectral shape on time scales ranging from months to a few minutes. Therefore, in order to shed light on the VHE emission mechanisms of blazars, simultaneous observations of these sources across multiple wavelengths are required. It is particularly important to study the connection between the variability between different wavebands. With the increased sensitivity of the latest generation of Imaging Atmospheric Cherenkov Telescopes (IACTs) such campaigns have become feasible for a larger number of sources. Three such campaigns were conducted in 2008: one on 1ES 1011+496 described in this contribution and the others on Mrk 180 and 1ES 2344+514 (see contribution 0832, R\"ugamer et al.).

1ES 1011+496 is an HBL located at a medium redshift of z=0.212. It was discovered at VHE by MAGIC in 2007 following an optical high state reported by the Tuorla Blazar Monitoring Programme\footnote{http://users.utu.fi/kani/1m} \cite{albert07}. Previous observations of this source at VHE showed only a hint of signal (see e.g. \cite{albert08a}) and the results presented here constitute the first follow-up detection of the source. The object shows frequent fluctuations in its optical brightness with the core flux increasing by up to a factor of $\sim$2-3 over the lowest states during flares. It has also been observed in the X-rays by Einstein \cite{elvis92} and the Swift/XRT \cite{abdo10} telescopes always showing a steep spectrum. There was no definite detection by EGRET \cite{sowards03} but it has been clearly detected by Fermi and was included already in the Fermi 3-month catalogue. The Fermi observations showed a very hard $\gamma$-ray spectrum  \cite{abdo10}. The MWL campaign undertaken in 2008 was the first of its kind on this source.

\section{Observation Campaign and Results}

\begin{figure}[t]
 \vspace{5mm}
 \centering
 \includegraphics[clip, width=1.\linewidth]{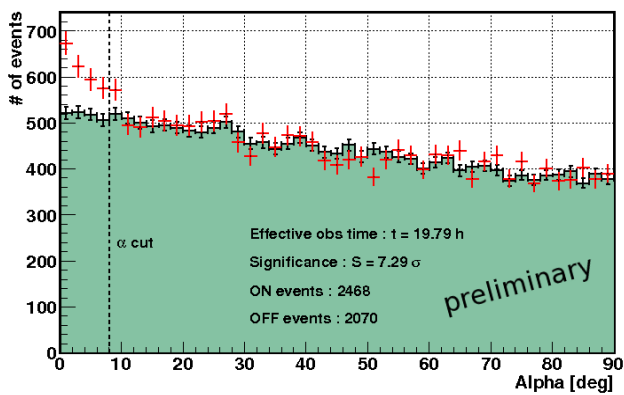}
 \caption{\small{Distribution of $\alpha$ angle for ON-region and OFF-region data. ON-data is marked by the red crosses while the filled region is the OFF-data. The signal region is marked by the dashed line.}}
 \label{det}
\end{figure}

MAGIC (Major Atmospheric Gamma-ray Imaging Cherenkov) is a system of two 17 m IACTs located on the Canary island of La Palma, Spain at $\sim$2200 m asl. At the time of the 2008 campaign the second telescope was still under construction and observations were made using only MAGIC-I which has been in operation since 2004. Thanks largely to its 236 m$^{2}$ reflective area MAGIC-I achieved an energy threshold of $\sim$60 GeV ($\sim$30 GeV with a special trigger) -- the lowest of any IACT. It reached a sensitivity of $\sim$1.6\% of Crab nebula flux (5 $\sigma$ in 50 h) $>$200 GeV with an energy resolution of $\sim$20-30\% and an angular resolution of $\sim$0.1$^{\circ}$ \cite{albert08b}.

The MWL observations were centered around common observation windows of the AGILE satellite \cite{tavani08} and the MAGIC-I telescope in spring 2008. Optical coverage was provided by the KVA telescope \cite{aleksic11} which is operated concurrently with the MAGIC telescopes. Observations in the radio were performed by the Mets\"ahovi radio telescope \cite{teräsranta98} and in the X-rays by the Swift satellite \cite{gehrels04}.

MAGIC was able to observe the object on 25 nights between March 4th and May 24th. The observation period was plagued by poor weather conditions at La Palma with frequent clouds, rainfall, strong wind and calima towards the end of the observation window. Nevertheless a total of 27.9 hours of data between zenith angles of 20$^\circ$ and 37$^\circ$ were collected of which 8 hours had to be removed due to adverse weather conditions. 20 hours of data were not or only marginally affected by bad weather and survived the quality cuts. The observations were done in wobble mode and the data analysed using the MAGIC standard analysis (for a comprehensive overview see e.g. \cite{albert08b}). The MAGIC observations resulted in the first confirmation of the source as a VHE emitter at 7 $\sigma$ significance level (see Figure \ref{det}).

\section{Discussion and Conclusions}

\begin{figure}[!b]
 \vspace{5mm}
 \centering
 \includegraphics[width=1.\linewidth]{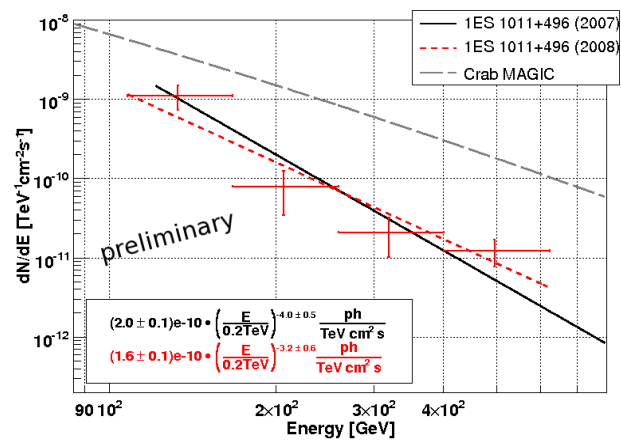}
 \caption{\small{The measured spectrum of 1ES 1011+496. The spectrum resulting from the 2008 observations is shown by the red crosses, with the dotted line representing a power law fit to the data. The black solid line gives the spectrum measured during the discovery. For comparison the Crab Nebula spectrum is shown (the grey dashed line).}}
 \label{spectrum}
\end{figure}

\begin{figure}[!t]
 \vspace{5mm}
 \centering
 \includegraphics[width=1.\linewidth]{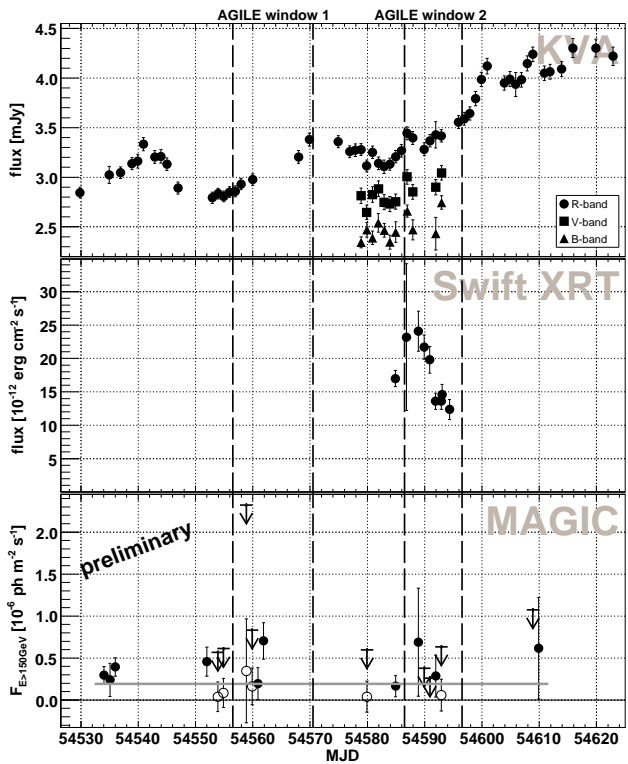}
 \caption{\small{The MWL light curve of 1ES 1011+496 with daily binning. The optical data (including the B and V band) are from the KVA telescope and are not host-galaxy corrected. In the MAGIC data, for points consistent with zero flux 99.7\% upper limits are shown and the grey horizontal line denotes a constant fit to the data points. The full circles are the significant data points while the open ones represent fluxes consistent with zero but with significance $>$0.}}
 \label{1011_MWL}
\end{figure}

MAGIC saw the source at a mean flux level similar to that measured during the discovery in 2007 \cite{albert07}. Also the derived spectrum was compatible within the error bars although a slight hint of a hardening of the spectral index in 2008 was seen (see Figure \ref{spectrum}). The VHE $\gamma$-ray spectrum can be described by the simple power law (with the differential flux given in units of TeV$^{-1}$m$^{-2}$s$^{-1}$):
\begin{eqnarray*}
\frac{\mathrm {d}N}{\mathrm{d}E} = (1.6\pm 0.1)\times10^{-10}\big(\frac{\mathrm E}{200\,\mathrm {GeV}}\big)^{-3.20\pm0.58}
\end{eqnarray*}
No significant variability of the VHE flux was detected during the campaign (see Figure \ref{1011_MWL}). A constant fit to the data yielded a $\chi^{2}$/d.o.f. of 13.9/18, corresponding to a probability of 73\%. Nevertheless in one night (MJD 54561.9) a flux has been found which is $>$2 sigma off the fit line.

Of the other telescopes Mets\"ahovi and AGILE did not detect the source within the given time windows. In the optical R band the source was at a relatively low state for most of the time with a major flare occurring at the end of the observation period. From the end of April to the beginning of May the source was also observed in the optical V and B band. The fluxes in these bands followed the same trends as in the R-band. In the X-rays a flare could be clearly detected as can be seen in the light curve in Figure \ref{1011_MWL}. The observation sampling prevented to define a baseline flux level, therefore the rise and fall times of the flare could not be evaluated. The Swift UVOT data are still being analysed.

\begin{figure}[!t]
 \vspace{5mm}
 \centering
 \includegraphics[width=1.\linewidth]{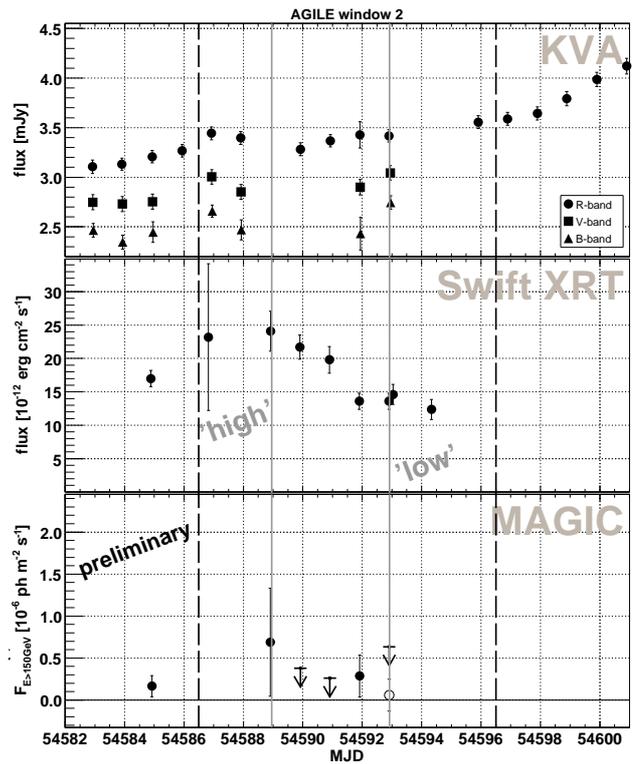}
 \caption{\small{A zoom into the MWL light curve around the Swift XRT flare. For an explanation of the symbols see Figure \ref{1011_MWL}. A 'high' and 'low' state has been defined according to the X-ray flux level which are used to compose simultaneous SEDs (see Figure \ref{1011_SED}).}}
 \label{1011_MWL_zoom}
\end{figure}

\begin{figure}[!b]
 \vspace{5mm}
 \centering
 \includegraphics[width=1.\linewidth]{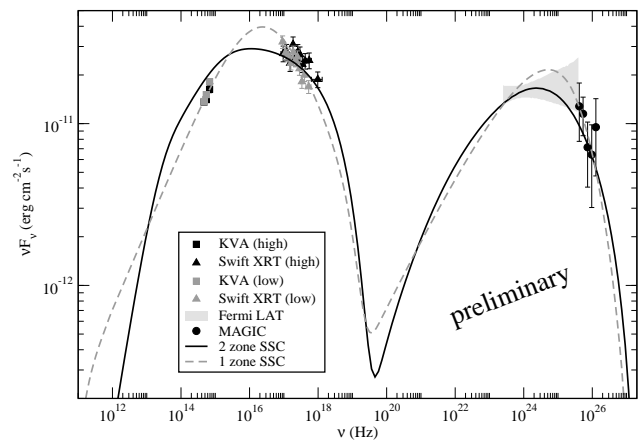}
 \caption{\small{MWL SEDs of 1ES 1011+496. The 'high' and 'low' data correspond to the observation times defined in Fig \ref{1011_MWL_zoom}. The KVA data are host-galaxy corrected following \cite{nilsson07}. The Fermi bow-tie deduced from the LAT 1-year Catalog \cite{abdo09} is added for illustrative purposes. The MAGIC data have been corrected for EBL absorption using \cite{kneiske10}. For a description of the model fits see text.}}
 \label{1011_SED}
\end{figure}

\begin{table*}
 \begin{center}
  \begin{tabular}{c|ccccccccc}
  \hline
  Model & B[G] & $\delta$ & R[cm] & K [cm$^{-3}$] & e$_{1}$ & e$_{2}$ & $\gamma_{min}$ & $\gamma_{break}$ & $\gamma_{max}$ \\
  \hline
  \cite{maraschi03} & 0.19 & 25 & 1.E+16 & 6.6E+03 & 2.0 & 4.1 & 1 & 4.E+04 & 6.E+05 \\
  \cite{weidinger10} & 0.18 & 44 & 8.E+15 & 7.5E+04 & 2.2 & 3.2 & 868 & 3.E+03 & 6.E+05 \\
  \hline
  \end{tabular}
 \caption{\small{Model parameters resulting from \cite{maraschi03} and \cite{weidinger10}. See text for details.}}
 \label{table_param}
 \end{center}
\end{table*}

There does not appear to be any obvious correlation between the optical and X-ray, X-ray and VHE or optical and VHE light curves (see Figure \ref{1011_MWL_zoom}). However the statistics of the individual data points of the VHE light curve are too low to draw definitive conclusions.

Two separate data sets, 'high' and 'low', defined according to the X-ray flux state of the source as can be seen in Figure \ref{1011_MWL_zoom}, were used to construct simultaneous SEDs (see Figure \ref{1011_SED}). Due to the small second observation window, no AGILE upper limit could be extracted for this period. The data has been modelled using a one-zone SSC \cite{maraschi03} as well as a self-consistent two-zone SSC model as described in \cite{weidinger10}. The small difference between the SEDs of the low and high X-ray flux states does not make a dedicated modelling of both seem meaningful. Therefore, only one model fit with each model was applied to the data for the time being.

Both models assume an emission region characterised by its radius R, magnetic field B and Doppler factor $\delta$ containing an electron population. The electrons with density K follow a broken power law distribution with index e$_{1}$ for $\gamma_{min}<\gamma<\gamma_{break}$ and e$_{2}$ for $\gamma_{break}<\gamma<\gamma_{max}$. In the case of \cite{weidinger10}, the electron distribution is characterised by their density K at $\gamma_{min}$ (note that for \cite{maraschi03}, K is defined at $\gamma$ = 1) with which they are injected into the acceleration region. The spectral indices and power law slopes are derived then self-consistently from cooling and acceleration processes.

The model parameters are given in table \ref{table_param}. The results represent standard values for HBLs except for the high Doppler factor and $\gamma_{min}$ of model \cite{weidinger10}. Further, more refined modelling is ongoing and necessary to confirm these results. Nevertheless both models fit the data rather well indicating that SSC mechanisms are the likely explanation for the observed emission.

Note that even though the Fermi bow-tie was not measured simultaneously to the other data it still fits well into the applied model scenario. Taking into account that also the VHE spectrum seems not to have changed considerably from the time of the discovery to these observations, 1ES 1011+496 seems to show a rather constant emission at $\gamma$-ray energies. At the lower energies, however, the optical flux measured by KVA is found at $\sim$30\% lower level and the X-ray flux at almost 10x higher level than the one derived from archival observations in the VHE discovery paper. The low (non-simultaneous) X-ray flux constraining the SED modelling led \cite{albert07} to the conclusion that in this source, the inverse-Compton component would dominate over the synchrotron component. On the contrary, the simultaneous SEDs from these observations indicate that this interpretation may not be true, corroborating that this source is synchrotron dominated like most HBLs.

In conclusion we organised the first MWL campaign including VHE coverage on the HBL 1ES 1011+496. The campaign resulted in the first confirmation of the source at VHE at a flux level similar to that measured during the discovery and a hint of hardening of the VHE spectrum compared to the discovery. The VHE flux was consistent with being constant, though significant variability in both the optical and X-ray wavelengths was detected. No apparent correlation between the flux levels in different wavebands was found. The data was modelled by a one-zone SSC and a self-consistent two-zone SSC model which fit the points well indicating that the SSC scenario is sufficient to describe the emission mechanisms of the source. However data analysis and more refined modelling is still ongoing. \\

{\renewcommand{\baselinestretch}{1.}\normalsize
\vspace{0.2cm}
\begin{scriptsize}
\textit{Acknowledgements:} The work of RR and EL has been supported by grant 127740 of the Academy of Finland. We would also like to thank the Instituto de Astrof\'{\i}sica de Canarias for the excellent working conditions at the Observatorio del Roque de los Muchachos in La Palma. The support of the German BMBF and MPG, the Italian INFN, the Swiss National Fund SNF, and the Spanish MICINN is gratefully acknowledged. This work was also supported by the Marie Curie program, by the CPAN CSD2007-00042 and MultiDark CSD2009-00064 projects of the Spanish Consolider-Ingenio 2010 programme, by grant DO02-353 of the Bulgarian NSF, by the YIP of the Helmholtz Gemeinschaft, by the DFG Cluster of Excellence "Origin and Structure of the Universe", by the DFG Collaborative Research Centers SFB823/C4 and SFB876/C3, and by the Polish MNiSzW grant 745/N-HESS-MAGIC/2010/0. The AGILE Mission is funded by the Italian Space Agency (ASI) with scientific and programmatic participation by the Italian Institute of Astrophysics (INAF) and the Italian Institute of Nuclear Physics (INFN).

We gratefully acknowledge N. Gehrels for approving this set of ToOs and the entire \textit{Swift} team, the duty scientists and science planners for the dedicated support, making these observations possible.

\end{scriptsize}}

\clearpage

\end{document}